\documentclass[12pt]{article}
\usepackage[cp1251]{inputenc}
\usepackage[russian]{babel}
\usepackage[T2A]{fontenc}
\begin{document}

\begin{center}{\bf On possibility of topological interpretation of quantum mechanics}\end{center}
\begin{center}{\bf O.A.Olkhov}\end{center}
\begin{center}{N.N.Semenov Institute of Chemical Physics, 119991 Moscow} \end{center}

\noindent E-mail: olega@gagarinclub.ru
\par\medskip
It is shown that the Dirac equation for free particles and Maxwell equations for free  electromagnetic waves can be considered as relations describing propagation of the topological defects of the physical three-dimensional space. These defects, being closed topological manifolds, can be considered as embedded in the outer five-dimensional space, and observable objects appear to be intersections of above defects on the physical space. This approach explains all irrational properties of quantum particles. Wave properties arise as a result of periodical movement of defect relative to its projection on physical space, and just this periodical movement attributes phase to the propagating particle. Appearance of probabilities within the formalism is a consequence of uncertainty of the closed topological manifold shape. Embedded in the outer space topological defects provide  channels  for nonlocal correlations between noninteracting particles in EPR-experiments. Mass, spin, light velocity happen to be topological invariants. It is  shown that the Dirac equation for hydrogen atom can be also considered as a relation describing the space topological defect. Electromagnetic potentials play here the role of connectivities of the universal covering space of the corresponding topological defect, and the gauge invariance of potentials is a natural consequence of the geometrical interpretation. Above result means that there is no need within topological approach to simulate hydrogen atom as a system with some point-like particles interacting trough their pair potentials. Acceptance  of suggested concept means rejection of the existing atomic paradigm based on the suggestion that matter consists of more and more small elementary particles. Within topological concept there are no any particles a priori,  before measurement procedure: notions of mass, 4-momentum, spin and of another attributes of particles appears only as a result of classical interpretation of the device reaction on its contact with the space defect representing quantum object. Suggested approach can be also considered as a nonlocal model with hidden variables.

\par\medskip
\begin{center}{\bf Outline}\end{center}
1.	Introduction\\
2.	What an electron is? Main idea\\
3.	Where the idea came from\\
4.	Propagation and probabilities\\
5.  Particles as topological defects. Wave-corpuscular duality\\
6.	Electromagnetic waves as topological defects. Light velocity invariance\\
7.	EPR-paradox\\
8.	Hydrogen atom without electron\\
9.	New paradigm. Perspectives
\par\medskip
\begin{center}{\bf 1.Introduction}\end{center}

In recent years considerable attention is given to the conceptual problems of quantum physics. One of these problems is a possibility to find out some reasonable physical model for quantum objects described with high efficiency by existing quantum formalism and to explain in so doing irrational properties of quantum world (probabilities, wave-corpuscular duality, EPR-paradox and so on) (see rewiews [1,2] and e.g. [3,4]). In short this problem is named as the problem of interpretation of quantum mechanics, and not long ago the majority of  specialists believed that there are no such problem at all: the opinion was established that Physics became too complicated to be described by physical models and that looking for models is useless as long as pure mathematics provides the progress. This opinion was supported by Dirac and even by Einstein (at the late period)[5,6,7]),though De Broglie never accepted this point of view [7].

Any way this problem was placed by Vitaly Ginsburg at first place between three "great" problems of modern physics (other two are "arrow of time" or irreversibility and phenomenon of life) [8].
May be, one of the reasons for the new attention to conceptual problems is the lack of new serious ideas that cause the feeling  " that something has to be done".( Indeed, the last such idea was Yang-Mills fifty years old idea about local gauge invariance). Other reason is probably the experimental approval of one of the most irrational properties of quantum world---instantaneous nonlocal correlation between noninteracting particles separated by macroscopic  distances  up to 10 km (paradox of Einstein, Podolsky and Rosen) (see, e.g.[9,10]). These experiments firstly directly indicated that definite states can not be attributed to microscopic systems as objective states independent of measurement procedure, and these investigations gave start to the new discipline-quantum information (see, e.g. [11,12]).

In this work we will represent topological model of quantum particles and electromagnetic waves, where these objects appear to be specific distortions of the Euclidean geometry of physical space-time. We will show below that such approach explains all irrational properties of quantum world--wave-corpuscular duality, probabilities, EPR-paradox, spin, invariance of light velocity and so on. There were many attempts to express formally notions of quantum formalism and electromagnetism  through geometrical notions, but only a few considered physical objects as deformations of space-time itself. The first was Einstein's theory of general relativity. Then, there were Wheeler's geometrodynamics [13] and, in a way, string theories (see, e.g. [14]). But Wheeler investigated possibility of topological interpretation of macroscopic electrical charges, whereas string theory works at Plank scales. Our approach works at atomic scales. Results, presented in this publication, were published portionwise in [15-20].
\par\medskip
\begin{center}{\bf 2. What an electron is? Main idea}\end{center}

In this Section we will represent pictorially (although rather schematic and conventional)the main idea of topological model for quantum particles. The reasonable foundations for the model will be considered in Sec.3.

A simple visual representation is possible only within one-dimensional model of the space because in three-dimensional space we have to use for explanation four-dimensional geometrical objects.
Let us consider one-dimensional world, where empty physical space is one-dimensional Euclidean space (straight 0X-axes). According to general relativity appearance of gravitational field leads to some curvature of this space--the space becomes the curved

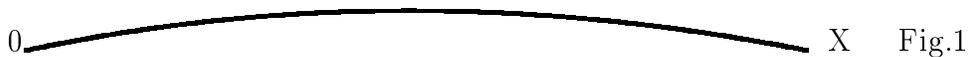
\begin{picture}(110,40)
{\linethickness{0,5mm}\qbezier(0,0)(142,30)(295,0)}
\put(303,0){X}
\put(-7,0){0}
\put(330,0){Fig.1}
\end{picture}
\par\medskip
\noindent Riemmann space (Fig.1). It was the first case in physics when some physical object was described not as something into the space, playing the role of a scene, but as a deformation of the space itself, and such deformation can be considered as the physical model for the gravitational field.

According to traditional interpretation appearance of an electron means appearance of "something" that moves in the space (Fig.2). The space plays here the role of a scene, and properties of this "something" are

\begin{picture}(110,80)
\put(0,0){\linethickness {0,5mm} \vector(1,0){295}}
\put(235,24){\vector(1,0){20}}
\put(20,20){$\exp\left(-\frac{i}{\hbar}(Et-px)\right)$}
\put(303,0){X}
\put(-7,0){0}
\put(135,20){"SOMETHING" ?}
\put(330,0){Fig.2}
\put(265,24){$\lambda=\frac{2 \pi \hbar}{p},\quad \omega=\frac{E}{\hbar}$}
\end{picture}
\par\medskip
\noindent described by the function that looks like a plane wave. The wave parameters of the object are defined through momentum and energy of the particle by the well known De Broglie's relations
$$\lambda=\frac{2 \pi \hbar}{p},\quad \omega=\frac{E}{\hbar}. \eqno (1)$$ There is no any physical model for an electron within this interpretation. Note also that relation (1) looks rather unnatural because notions of more general microscopic wave theory ($\lambda,\omega$) are expressed there through notions of less general macroscopic classical theory (p,E).(For example, more general relativistic classical mechanics can be formulated without using less general Newtonian mechanics).

Within suggested concept, appearance of an electron in the empty one-dimensional space means appearance of topological defect of this space. This defect can be shown (rather schematic) as an one-dimensional branch of the physical space, and this branch can be considered as the piece of one-dimensional space embedded in the outer three-dimensional space (the space topological defects are the space deformations that can not be eliminated by continuous transformations). The observable object (electron ) is an intersection of the topological defect with one-dimensional physical space (Fig.3.)(In this simple example the intersection is a geometrical point, but we will show later that in three-dimensional space such intersection looks like  a region of the space).

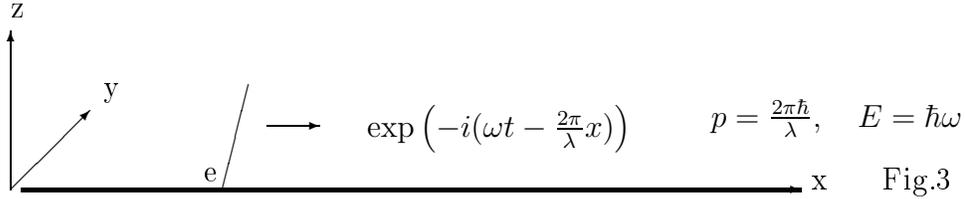
\begin{picture}(110,80)
\put(0,0){\linethickness {0,5mm} \vector(1,0){295}}
\put(0,0){\vector(0,1){60}} \put(0,0){\vector(1,1){30}}
\put(0,65){z}\put(35,35){y}
\put(80,0){\line(1,4){10}} \put(97,24){\vector(1,0){20}}
\put(303,0){x}
\put(135,20){$\exp\left(-i(\omega t-\frac{2 \pi}{\lambda}x)\right)$}
\put(265,24){$p=\frac{2 \pi \hbar}{\lambda},\quad E=\hbar \omega$}
\put(330,0){Fig.3}
\put(73,3){e}
\end{picture}

The above topological defect (and its intersection point) propagates along the one-dimensional physical space performing simultaneously periodical movement in the outer three-dimensional space, and within suggested model the wave function
$\exp\left(-i(\omega t-2 \pi x/\lambda)\right)$ describes just this periodical process. Within this model classical notions of momentum and energy for above topological object appear only as a result of interpretation on classical language of the effects produced into macroscopic device by its contact with the piece of deformed space. Above classical notions are connected with microscopic ones by relations
$$p=\frac{2 \pi \hbar}{\lambda},\quad E=\hbar \omega.\eqno(2)$$
These relations formally coincide with the De Broglie definition, but they look more natural because
they express the notions of less general theory (classical notions of momentum and energy) through the notions of more general microscopic theory (parameters of periodical movement of the space topological defect). Note once more that within suggested topological model we obtain simple explanation of the wave properties of the point-like particle: being the intersection point with the periodically moving defect such particle also transfer the phase of this movement.
\par\medskip
\begin{center}{\bf 3.Where the idea came from}\end{center}

In previous Section we schematically presented the model of quantum particles as some topological objects, and we presented this model without any approvement. But, there can not be rigorous approvement for new concept: there can be only reasonable arguments and confirmation by experiment. In this Section we will demonstrate these arguments, and we will show that suggested topological model explains all irrational observable properties of quantum objects .

Considering a possibility of geometrization of quantum physics, we have to keep in mind the exceptional accuracy of the modern quantum formalism. Therefore, we suppose that attempts to find out new geometrical description of quantum objects have to begin not with the creation of a new mathematical formalism but with finding out a geometrical interpretation of the well-known basic relativistic quantum
equations whose validity is beyond question. We considered three such equations:the Dirac equation for free particle, the Maxwell equations for free electro-magnetic field and the Dirac equation for hydrogen atom.

We start at first with geometrical interpretation of the Dirac equation for free particle (for free matter wave field with spin $1/2$). This equation can be written in the following form (see, e.g.,[21])
$$i\gamma^{\mu}\partial_{\mu}\psi=m\psi,\eqno (3)$$
where $\partial_{\mu}=\partial/\partial x_{\mu},\quad \mu
=1,2,3,4$,\quad $\psi (x)$ is the four-component Dirac bispinor,
$x_{1}=t, x_{2}=x, x_{3}=y$, $x_{4}=z$, and $\gamma^{\mu}$ are
four-row Dirac matrices (concrete representations of $\gamma$-matices and bispinor are of no importance). The summation in Eq.(3) goes over the
repeating indices with a signature $(1,-1,-1,-1)$. Here, $\hbar
=c=1$, $m$ is the particle mass. For definite values of 4-momentum $p_{\mu}$, the solution to
Eq.(1) has a form of the plane wave
$$\psi=u(p_{\mu})\exp (-ip_{\mu}x^{\mu}),\eqno(4)$$ where $u(p_{\mu})$ is a normalized bispinor.
Substitution of (2) in Eq.(1) gives the known relation for
$p_{\mu}$
$$p_{1}^{2}-p_{2}^{2}-p_{3}^{2}-p_{4}^{2}=m^{2}.\eqno(5)$$

Eq.(3) serves as a basis for describing of experiments with fine
accuracy in the intermediate energy range where relativistic
corrections can not be ignored but where we can neglect of the
possibility of new particles appearance (for describing fine structure of the hydrogen atom spectra, for example). Wave function (4) is
interpreted as a description of some free wave field with mass $m$
and spin $1/2$. The wave function amplitude squared defines the
probability to find particle in the corresponding point and the
phase factor in (4) describes the wave--corpuscular properties of a
particle using postulated relations between particle's 4-momentum
$p_{\mu}$ and "particle's wavelength" $\lambda_{\mu}$
$$\lambda_{\mu}=2\pi p_{\mu}^{-1}.\eqno (6)$$ Any model or any pictorial
representation of the quantum object is absent, and the space-time
is considered as a scene where such objects exist and interact.
Having in mind geometrical interpretation of Eq.(3), we rewrite
function (4) and relation (5) in the form where only geometrical notions with dimensionality of length are presented
$$\psi=u(p_{\mu})\exp (-2\pi ix^{\mu}\lambda_{\mu}^{-1}).\eqno(7)$$
$$\lambda_{1}^{-2}-\lambda_{2}^{-2}-\lambda_{3}^{-2}-\lambda_{4}^{-2}=\lambda_{m}^{-2}, \quad \lambda_{m}=2\pi m^{-1}.\eqno(8)$$

The first idea looks as follows: if Eq.(3) has some geometrical meaning then there should be some reasons for using in this equation the specific tensors---spinors. It is known that
there is a correspondence between every kind of tensors and some
class of geometrical objects in the sense that these tensors define
invariant properties of above objects. For example, usual vectors
correspond to simplest geometrical objects--to points [22], and this
is one of reasons why Newtonian mechanics uses vectors within its
formalism. Spinors correspond to nonorientable geometrical object
(see, e.g., [23]). So, we suppose that spinors are used in Eg.(3),
because this equation describes some nonorientable geometrical
object and "$spin = 1/2$" is a formal expression of the
nonorientable property of the object.

The above assumption is a starting idea.
To define properties of the
proposed geometrical object more exactly we consider more precisely
the symmetry properties of the solution (7). First of all, function (7) is an
invariant with respect to coordinates transformations
$$x^{'}_{\mu}=x_{\mu}+n_{\mu}\lambda_{\mu}, \quad n_{\mu}=0,\pm 1, \pm 2, ...\quad .\eqno (9)$$
Transformations (9) can be considered as elements of the group of
translations operating in the Minkovsky 4-space where wave function (7) is
defined. Then function (7) can be considered as a vector realizing
this group representation. From the other hand, as a bispinor, function (7) realizes representation of one more
group of the symmetry transformation of 4-space that is not so
obvious. Being a four-component spinor, $\psi(x)$ is related to the
matrices $\gamma^{\mu}$ by the equations (see, e.g. [24])
$$\psi^{'}(x^{'})=\gamma^{\mu}\psi (x),$$ where $x\equiv
(x_1,x_2,x_3,x_4)$, and $x^{'}\equiv (x_1,-x_2,-x_3,-x_4)$ for $\mu
=1,x^{'}\equiv (-x_1,x_2,-x_3,-x_4)$ for $\mu =2$, and so on. This
means that the matrices  $\gamma^{\mu}$ are the matrix
representation of the group of reflections along three axes
perpendicular to the $x_\mu$ axis, and the Dirac bispinors realize
this representation.

Above two groups form a group of four sliding symmetries with
perpendicular axes (sliding symmetry means translations plus
corresponding reflections; see e.g.,[25]). The physical Minkovsky space-time does not
have such symmetry. But this group may operate in some auxiliary
space: it is known within topology that discrete groups
operating in some auxiliary space can reflect a symmetry of geometrical
objects that have nothing in common with this space. It will be the
case when such space is a universal covering space of some closed
topological manifold. Universal covering spaces are auxiliary spaces
that are used in topology for the description of closed manifolds,
because discrete groups operating in these spaces are isomorphic to
fundamental groups of closed manifolds---groups whose elements are
different classes of closed pathes on manifolds started and finished at the same point (so called $\pi_{1}$
group [25-27]). For example, fundamental group of the nonorientable two-dimensional Klein bootle is a group of two sliding symmetries with parallel axes, and the universal covering space of this closed manifold is a two-dimensional Euclidean space [25,26].
Above properties of solution (7) leads to the idea that Dirac Eq.(3) describes, in fact, some closed
nonorientable space-time 4-manifold, whoose fundamental group is a group of the above four sliding symmetries, and space-time plays
not only the role of the "space of events" but it plays also the role of the universal covering space for above manifold.

To clarify the possibility of "coding" of some closed manifold by means of linear differential equation  we consider below the simplest example of closed topological manifold--manifold homeomorphic  ("equivalent") to a circle with given perimeter length $\lambda=2\pi R$. Such manifold is represented by any of its deformation without discontinuities and gluing (Fig.4).

\begin{picture}(110,50)
\put(5,25){\circle{25}} \put(55,25){\oval(15,30)}
\put(110,25){\oval(40,10)}
\qbezier(160,35)(215,25)(160,20)\qbezier(160,35)(240,25)(160,20)\qbezier(225,25)(230,40)(235,30)
\qbezier(225,25)(225,30)(230,15)\qbezier(235,30)(240,35)(250,25)\qbezier(230,15)(235,27)(250,25)
\put(330,25){Fig.4}
\end{picture}

The fundamental group of above manifold is a group isomorphic to the group of integers [26,27]. This is illustrated at Fig.5: all possible closed pathes on any of representative of this manifold differ one from another by number of circuits started at the same point of manifold, and any of such pathes can be considered as a result of summation of another pathes.

\begin{picture}(110,67)
\put(37,35){\oval(30,30)}
\qbezier(25,35)(30,50)(35,40)
\qbezier(25,35)(25,40)(30,25)\qbezier(35,40)(40,45)(50,35)\qbezier(30,25)(35,37)(50,35)
\put (20,0){1 circuit}
\put(137,35){\oval(30,30)}
\qbezier(125,35)(130,50)(135,40)
\qbezier(125,35)(125,40)(130,25)\qbezier(135,40)(140,45)(150,35)\qbezier(130,25)(135,37)(150,35)
\put (120,0){2 circuits}
\put(137,35){\oval(35,35)}
\put(237,35){\oval(30,30)}
\qbezier(225,35)(230,50)(235,40)
\qbezier(225,35)(225,40)(230,25)\qbezier(235,40)(240,45)(250,35)\qbezier(230,25)(235,37)(250,35)
\put (220,0){3 circuits}
\put(237,35){\oval(35,35)}
\put(237,35){\oval(40,40)}
\put(88,0){$+$}
\put(188,0){$=$}
\put(330,25){Fig.5}
\end{picture}
\par\medskip

From the other hand, the group of integers isomorphic to the group of translation operating in the one-dimensional Euclidean space, and just this space plays the role of universal covering space of our manifold (Fig.6).

\begin{picture}(110,60)
\put(0,20){\linethickness {0,5mm} \vector(1,0){295}}
\put(50,20){\circle*{5}}
\put(100,20){\circle*{5}}
\put(150,20){\circle*{5}}
\put(200,20){\circle*{5}}
\put(250,20){\circle*{5}}
\put(48,5){$1$}
\put(98,5){$2$}
\put(148,5){$3$}
\put(198,5){$4$}
\put(248,5){$5$}
\put(25,30){$\lambda$}
\put(75,30){$\lambda$}
\put(125,30){$\lambda$}
\put(175,30){$\lambda$}
\put(225,30){$\lambda$}
\put(305,20){$X$}
\put(330,20){Fig.6}
\end{picture}

The group of integers isomorphic to the group of translations that is represented by one-dimensional matrixes $\exp (-2\pi i n)$ with the basic vector $f(x)=f(x+\lambda)$
$$f(x)=A\exp (2 \pi i x/\lambda). \eqno(10)$$ This function is an analog of the Dirac equation solution (7) and this function itself is a solution of the equation
$$i\frac{\partial f}{\partial x}=mf,\quad m=\frac{2\pi}{\lambda},\eqno(11)$$and this equation can be considered as an analog of the Dirac Eq.(3). So, above consideration shows how linear differential equation can contain information about symmetry of closed manifolds, although Eq.(11) can, of course, describe many other phenomena.

Finally, the main hypothesis looks as follows: {\it the Dirac Eq.(3)(and consequently its solution (7)) describes some specific deformation of the space-time itself, namely, the closed nonorientable topological space-time 4-manifold whose fundamental group is a group of four sliding symmetries with perpendicular axes. Spin $1/2$ is a topological invariant reflecting nonorientable character of the manifold. Mass $m$ and 4-momentum $p_{\mu}$ are classical notions defined through the geometrical (microscopical) parameters of the manifold by relations (2,8) and Planck constant $\hbar$ and the light velocity $c$ play the role of coefficients of transfer from one system of notion to other.}

At the present time, only two-dimensional Euclidean closed
manifolds are classified in details, and their fundamental groups
and universal covering planes are identified [26,27]. As it is known to
author, four-dimensional manifolds with above fundamental group
operating in pseudoeuclidean universal covering space were not
considered before. Therefore, we have
no opportunity for rigorous consideration of specific properties of
suggested pseudoeuclidean geometrical object. But qualitative properties,
explaining main ideas of new interpretation, can be investigated
using low-dimensional analogies. Using these analogies we will show
within elementary topology that the above 4-manifold represents
propagation of the topological defect of three-dimensional euclidean
space in a way, described in Sec.2, and that propagation of this defect demonstrates specific
properties of quantum particles: stochastic behavior, wave-corpuscular dualism and quantum nonlocality.
\par\medskip
\begin{center} {\bf 4.Propagation and probabilities} \end{center}

Let us consider the simplest example of a
closed topological manifold---one-dimensional manifold homeomorphic
to a circle whose perimeter length is fixed and equals $\lambda$ (see Fig.4).
A closed topological manifold is representable by any of its
possible deformations (without pasting) that conserve manifold's
continuity, and we will see that just this "stochastic" property of topological manifolds explains
appearance of probabilities in quantum formalism. For simplicity we
consider only plane deformations of the circle.

To use concrete calculations, we consider manifold's
deformations that have a shape of ellipse with perimeter length $\lambda$ (Fig.7).

\begin{picture}(110,100)
\put(5,70){\circle{25}} \put(55,70){\oval(15,30)}
\put(110,70){\oval(40,10)}
\qbezier(160,80)(215,70)(160,65)\qbezier(160,80)(240,70)(160,65)\qbezier(225,70)(230,85)(235,75)
\qbezier(225,70)(225,75)(230,60)\qbezier(235,75)(240,80)(250,70)\qbezier(230,60)(235,72)(250,70)
\put(55,15){\oval(30,45)}
\put(55,15){\vector(1,0){30}}
\put(55,15){\vector(0,1){32}}
\put(58,45){$y$}
\put(81,18){$x$}
\put(59,7){$a_{1}$}
\put(45,20){$b_{1}$}
\put(175,15){\oval(45,30)}
\put(175,15){\vector(1,0){32}}
\put(175,15){\vector(0,1){30}}
\put(178,45){$y$}
\put(201,18){$x$}
\put(179,7){$a_{2}$}
\put(165,20){$b_{2}$}
\put(25,70){\circle*{2}}
\put(35,70){\circle*{2}}
\put(70,70){\circle*{2}}
\put(80,70){\circle*{2}}
\put(152,70){\circle*{2}}
\put(142,70){\circle*{2}}
\put(206,70){\circle*{2}}
\put(216,70){\circle*{2}}
\put(258,70){\circle*{2}}
\put(268,70){\circle*{2}}
\put(278,70){\circle*{2}}
\put(6,15){\circle*{2}}
\put(16,15){\circle*{2}}
\put(26,15){\circle*{2}}
\put(103,15){\circle*{2}}
\put(113,15){\circle*{2}}
\put(123,15){\circle*{2}}
\put(133,15){\circle*{2}}
\put(220,15){\circle*{2}}
\put(230,15){\circle*{2}}
\put(240,15){\circle*{2}}
\put(330,15){Fig.7}
\end{picture}\

The equation for the ellipse on an euclidean plane has
the form
$$x^2/a^2+y^2/b^2=1, \eqno (12)$$where all possible values of the
semiaxes $a$ and $b$ are connected with the perimeter length
$\lambda$ by the known approximate relation
$$\lambda\simeq \pi[1,5(a+b)-(ab)^{1/2}].\eqno (13)$$This means that the
range of all possible values of $a$ is defined by the inequality
$a_{min}\leq a\leq a_{max}\simeq \lambda/1,5\pi, a_{min}\ll
a_{max},$.

In the pseudoeuclidean "space-time," the equation
for our ellipses has the form of equation for hyperbola (after substitution $y=it$)
$$x^2/a^2-t^2/b^2=1, \eqno (14)$$ and this equation defines the
dependence on time $t$ for a position of the point $x$ of the
manifold corresponding to definite $a$. At $t=0,x=\pm a$; that is,
our manifold is represented by the two point sets in one-dimensional
euclidean space, and the dimensions of these point sets are defined
by all possible values of $a$. So, at $t=0$, the manifold is
represented by two regions of the one--dimensional euclidean space
$a_{min}\leq| x|=a\leq a_{max}$. It can easily be shown that at
$t\neq 0$ these regions increase and move along the x-axis in
opposite directions (at Fig.8 this movement is shown only for
positive direction).
\begin{center}
\begin{picture}(250,40)
\put(0,25){\vector(1,0){230}}
\put(5,25){\linethickness{1.0mm}\line(1,0){30}}\put(55,25){\linethickness{1.0mm}\line(1,0){50}}
\put(125,25){\linethickness{1.0mm}\line(1,0){70}}\thinlines{\put(35,35){\vector(1,0){18}}
\put(105,35){\vector(1,0){18}}\put(195,35){\vector(1,0){18}}\put(235,25){x}}
\put(20,35){$t_0$}\put(80,35){$t_1$}\put(160,35){$t_2$}
\put(290,25){Fig.8}

\end{picture}
\end{center}

 All another possible deformations of our circle will be obviously
represented by points of the same region, and every such point can
be considered as a possible position of the "quantum object"
described by our manifold. All manifold's deformations are realized
with equal probabilities (there are no reasons for another
suggestion). Therefore, all possible positions of the point--like
object into the region are realized with equal probabilities. So, this
example shows the possibility of the consideration of above object
as a point with probability description of its positions as it
suggested within standard representation of quantum particles. In
fact, this point is not yet a material point---it is a geometrical
point only. In the next Section we will show how this point becomes
the material point.

It should be stressed that within suggested approach stochastic behavior is a property of a single  quantum particle: the role of statistical ensemble plays here the ensemble of all possible topological realizations of the same particle---ensemble of all possible deformations of the same topological manifold (these deformations can be considered as some kind of hidden variables).

\begin{center}{\bf 5.Particle as topological defect. Wave-corpuscular duality}\end{center}

The simple example of preceding Section does not explain what
geometrical properties allows to differ points of the moving region
from neighbour points of the euclidean space making them observable.
To answer at this question let us consider more complex analogy of
the closed 4-manifold---two-dimensional torus. In euqlidean 3-space
such torus is denoted in topology as a production of two
one-dimensional closed manifolds $S^1\times S^1$ . The role of
different manifold's deformations as a reason for stochastic
behaviour was considered in preceding Section. Therefore, now we
restrict our consideration to one simplest configuration when one of the
$S^1$ is a circle in the plane $XY$ and another is a circle in the
plane $ZX$ (we denote it as $S_{t}^1$, see Fig.9).

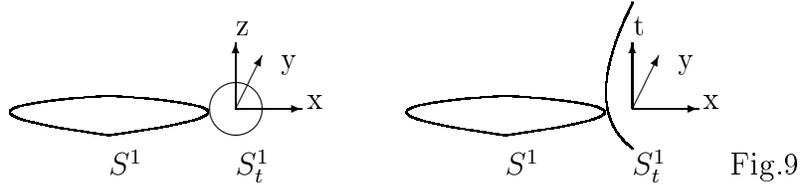
\begin{picture}(110,100)
\qbezier(100,10)(175,20)(100,25)
\qbezier(100,10)(25,20)(100,25)
\put(148,20){\circle{20}}
\put(148,20){\vector(1,0){25}}
\put(148,20){\vector(0,1){25}}
\put(148,20){\vector(1,2){10}}
\put(175,20){x}
\put(148,48){z}
\put(165,35){y}
\put(100,-5){$S^{1}$}
\put(148,-5){$S^{1}_{t}$}
\qbezier(250,10)(325,20)(250,25)
\qbezier(250,10)(175,20)(250,25)
\put(250,-5){$S^{1}$}
\put(298,-5){$S^{1}_{t}$}
\qbezier(298,5)(278,20)(298,60)
\put(298,20){\vector(1,0){25}}
\put(298,48){t}
\put(298,20){\vector(0,1){25}}
\put(325,20){x}
\put(298,20){\vector(1,2){10}}
\put(315,35){y}
\put(335,-5){Fig.9}
\end{picture}\
\par\medskip

In pseudueuclidean space this torus looks like a hyperboloid.
The
hyperboloid appears if we replace the circle $S_{t}^1$ by a
hyperbola (as it was done in Section 2). Positions of the
geometrical object described by our pseudoeucliden torus are defined
by time cross-sections of the hyperboloid. These positions looks
like as an expanding circle into two-dimensional xy-euclidean plane (Fig.10).
\par\bigskip
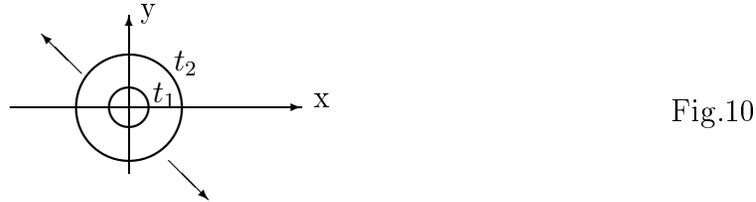
\begin{center}
\begin{picture}(250,60)
\put(40,45){\vector(1,0){110}}
\put(85,20){\vector(0,1){60}}\thicklines{
\put(85,45){\circle{50}}\put(85,45){\circle{15}}}\thinlines{\put(94,47){$t_1$}
\put(102,59){$t_2$}\put(155,45){x}\put(89,79){y}\put(67,58){\vector(-1,1){15}}\put(100,25){\vector(1,-1){15}}}
\put(290,40){Fig.10}
\end{picture}
\end{center}
But we need to have in mind that two-dimensional pseudoeuclidean
torus describes the object existing into two-dimensional space-time
with one-dimensional euclidean "physical" space. This means that an
observable part of the object is represented in our example by the
points of intersections of above circle with $0X$ axis though, as a
whole, the object is represented by a circle "embedded" into
two-dimensional, "external" space. This circle can be considered as
a topological defect of the physical one-dimensional euclidean
space (in Sec.2 we schematically represented such defect as a branch of the one-dimensional space). Just an affiliation of the intersection points to the
topological object differs these points geometrically from
neighboring points of the one-dimensional euclidean space.
Therefore, in pseudoeuclidean four-dimensional physical space-time
the suggested object described by the Dirac equation looks like a
topological defect of physical euclidean 3-space that is embedded
into 5-dimensional euclidean space, and its intersection with
physical space represents an observable quantum object.

Note, that expanding circles at Fig.10 can be considered as a model for propagation in opposite directions  of two identical noninteracting particles. Being the intersection points of the same defect with the physical space these particles can correlate one with another without any interaction in physical space--the channel for information is provided by their common defect embedded in the outer space. This can be considered as a qualitative explanation for the paradox of Einstein, Podolsky, Rosen (in Sec.7 we will consider this problem with more details).

Above analogy with torus does not yet demonstrate appearance of any
wave-corpuscular properties of the object, represented in "physical"
one-dimensional space by the moving intersection point--- properties
that could be expressed by wave function (5) and relation (4). In
the case of considered two-dimensional "space--time" this solution
has the form
$$\psi=u(p)\exp (-2\pi ix^{1}\lambda_{1}^{-1}+2\pi ix^{2}\lambda_{2}).\eqno(15)$$
Topological defect represented by the expanding circle does not
demonstrate any periodicity when the intersection point (physical
object) moves along one-dimensional euclidean $0X$--space. To demonstrate such periodicity we have to use now more correct low-dimensional analogy---some nonorientable one (torus is an orientable manifold).

The nonorientable Klein bottle could be
such two--dimensional analogy [25,26]. In the case with torus
topological defect was represented by cross-sections of
pseudoeuclidean torus--plane circles. The Klein bootle is a manifold
that is obtained by gluing of two Mobius strips (see, e.g.[28]).
Therefore, the Klein bootle cross-section is an edge of the Mobius
strip. This edge can not be placed in the two--dimensional
$XY$--plane without intersections, and it means that corresponding
topological defect is now a closed curve embedded into
outer three--dimensional $XYZ$--space (this does not give an opportunity to represent such defect at the plane figure as at Fig.8 for torus).
In this case the position of the topological defect (closed curve)
relative to its intersection with $0X$ axis (physical object) can
change periodically. The parameters of this periodical movement
depends on geometrical parameters $\lambda_{1}$ and $\lambda_{2}$
(there are no other parameters with corresponding dimensionality).
Such periodical process can be expressed by the function (15). This
gives an opportunity for the new interpretation of the wave function
as a description of periodical movement of the topological defect
relative to its projection on the physical space.

In this case corpuscular properties of the above periodical movement
appear as a result of the definition for classical notion of
4-momentum trough the wave characteristic of the topological object,
namely
$$p_{\mu}=2\pi/\lambda_{\mu}.\eqno (16)$$
Substitution of these relations into (15) leads to the Dirac
solution (4)
$$\psi=u(p)\exp (-ip_{1}x^{1}+ip_{2}x^{2}).\eqno(17)$$
It is important to note that within suggested geometrical
interpretation the notions of the less general, macroscopic theory
(4-momentums) are defined by (16) trough the notions of more general
microscopic theory (wave parameters of the defect periodical
movement), and this looks more natural than the opposite
definitions (6) within traditional interpretation.

\begin{center}{\bf 6.Electromagnetic waves as topological defects. Light velocity invariance}\end{center}

Let us write the Maxwell equations for electromagnetic waves in
vacuum in the symbolic form analogous to the Dirac equation (3).
Namely, we write these equations in the Majorana form [29]
$$i\frac {\partial \textbf{f}^+}{\partial t}=(\textbf{S} \textbf{p}) \textbf{f}^+, \quad \textbf{p} \textbf{f}^+ =0,$$
$$i\frac {\partial \textbf{f}^-}{\partial t}=-(\textbf{S} \textbf{p}) \textbf{f}^-, \quad \textbf{p} \textbf{f}^-
=0,\eqno (18) $$ where
$$\textbf{f}^+=\textbf{E}+i\textbf{H},\quad
\textbf{f}^-=\textbf{E}-i\textbf{H}.\eqno (19)$$ Here $\textbf{E}$
is an electric field, $\textbf{H}$ is a magnetic field,
$\textbf{p}=-i\nabla$ and $\textbf{S}$ is a vector-matrix
$$
S_{x}=\left(\begin{array}{ccc}0&0&0\\0&0&-i\\0&i&0\end{array}\right),
\quad
S_{y}=\left(\begin{array}{ccc}0&0&i\\0&0&0\\-i&0&0\end{array}\right),
\quad
S_{z}=\left(\begin{array}{ccc}0&-i&0\\i&0&0\\0&0&0\end{array}\right).\eqno
(20)
$$

It can be easily seen that Egs.(18) may be rewritten in the symbolic
form analogous to the symbolic form of the Dirac equation (1)
$$i\Gamma^{\mu} \partial_{\mu} \textbf{f}=0,\eqno (21)$$ where
bivector \textbf{f} and matrix $\Gamma^{\mu}$ have the form
$$
\Gamma^0=\left(\begin{array}{cc}0&1\\1&0\end{array}\right), \quad
\Gamma^{1,2,3}= \left(\begin{array}{cc}0&-{\bf S }\\{\bf
S}&0\end{array} \right),\quad
\textbf{f}=\left(\begin{array}{cc}\textbf{f}^+\\\textbf{f}^-\end{array}\right).
$$ We write here six-row matrices trough three-row ones. We see
that Egs.(18) has formally the same form as the Dirac equation (3)
with $m=0$. Only instead of the Dirac bisinor $\psi$ we have here
bivector \textbf{f}, and instead Dirac matrices $\gamma^{\mu}$ we
have matrices $\Gamma^{\mu}$. As Maxwell's Egs.(21) looks formally
like Dirac's Egs.(3) it seems reasonable to use for their
geometrization the same arguments as we used for the Dirac equation
geometrization.

For plane waves the solution of Eq.(21) has the form
$$\textbf{f}^+=\textbf{f}^+_k\exp i(\textbf{kr}-\omega t),\quad \textbf{f}^-=\textbf{f}^-_k\exp i(\textbf{kr}-\omega t),\quad \omega = |\textbf{k}|.
\eqno (22)$$ As for topological interpretation of solution (7) of
the Dirac equation we suggest that function (22) describes periodical movement of the
space topological defect. As function (7) solution (22) can be also
considered as the realization of fundamental group of some closed
topological manifold. Due to exponential factor in (22) this group
contains the translation group (as for solution (7)), but now our
solution are complex bivectors--not a bispinors. These vectors does
not realize the representation of reflections along three different
axes as it was for bispinors. These vectors $\textbf{f}^+$ and
$\textbf{f}^-$ consist of axial and polar vectors (see (19)) and
thus these vectors are transformed one into another only in result
of the reflection of space axes. Therefore, solution (22) realizes
representation of a sliding symmetry in 4-space only along
time--axis. This distinguishes the supposed fundamental group from
the fundamental group considered in previous Sections (four sliding
symmetries along four perpendicular axes).

There is no investigation in topology, where 4-manifolds with above
fundamental group were considered. So, we again can establish
connections between geometrical properties of the manifold and
observable physical properties of electromagnetic waves only using
low-dimensional analogies. Wave-corpuscular dualism of
electromagnetic waves and possibility of stochastic behavior can be
demonstrated in the same manner as in Sections 4,5. But
electromagnetic waves have some additional important
property---their velocity does not depend on the source motion. We
will show below how geometrical properties of the closed topological
4-manifold can explain this fact.

Suppose that 4-manifold corresponding to electromagnetic wave has
the form $M^3(\textbf{r})\times M^1(\textbf{r},t)$, that is it can
be represented as a product of nonorientable three-dimensional
euclidean closed manifold $M^3(\textbf{r})$ and one-dimensional
manifold $M^1(\textbf{r},t)$ homeomorphic to a pseudoeuclidean
circle. The formal reason for such representation is the
distinguished role of the euclidean space within fundamental group:
only into euclidean subspace translation group is combined with
reflections. Consider now a low-dimensional analogy that explains an
independence of light velocity on the source motion.

Instead of four-dimensional manifold $M^3(\textbf{r})\times
M^1(\textbf{r},t)$ we consider, as in previous Section,
two-dimensional analogy---manifold $S^1\times S_1'$, where $S^1$ is
a one-dimensional euclidean circle and $S_1'$ is a pseudoeuclidean
circle. This manifold was considered in Section 5, and it looks like
a hyperboloid. For electromagnetic waves $m=0,E=cp$. Within our
notation it leads to relation $p_{1}=p_{2}=p$ or
$\lambda_{1}=\lambda_{2}=\lambda$. Therefore, there have to be only
one parameter with dimensionality of length, and this will be the
case if $S_t^1$ is a pseudoeuclidean circle of zeroth radius.
Equation for such circle has the form $x^{2}-t^{2}=0$, and
hyperboloid is transformed into a cone (Fig.11).

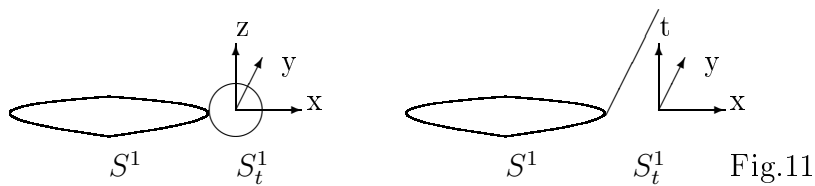
\begin{picture}(110,100)
\qbezier(100,10)(175,20)(100,25)
\qbezier(100,10)(25,20)(100,25)
\put(148,20){\circle{20}}
\put(148,20){\vector(1,0){25}}
\put(148,20){\vector(0,1){25}}
\put(148,20){\vector(1,2){10}}
\put(175,20){x}
\put(148,48){z}
\put(165,35){y}
\put(100,-5){$S^{1}$}
\put(148,-5){$S^{1}_{t}$}
\qbezier(250,10)(325,20)(250,25)
\qbezier(250,10)(175,20)(250,25)
\put(250,-5){$S^{1}$}
\put(298,-5){$S^{1}_{t}$}

\put(308,20){\vector(1,0){25}}
\put(308,48){t}
\put(308,20){\vector(0,1){25}}
\put(335,20){x}
\put(308,20){\vector(1,2){10}}
\put(325,35){y}
\put(335,-5){Fig.11}
\put(288,18){\line(1,2){20}}
\end{picture}\
\par\bigskip
This means that the
points representing in this example electromagnetic wave move with
velocity equals $\pm1$ ($\pm c$--in chosen units system), and this
velocity does not depend on coordinate frame rotations (does not
depend on transfer from one moving inertial frame to another). From
topological point of view this result is a consequence of the fact
that the zeroth radius can be considered as topological invariant.
Therefore, within geometrical approach light velocity appears to be
topological invariant of the manifold representing electromagnetic
wave, and this is the reason of its independence of the source
motion.
\par\medskip

\begin{center}{\bf 7. EPR-paradox}\end{center}

The most strange and irrational property of the quantum world is the so-called quantum nonlocality, when physical states of noninteracting particles, separated by macroscopic distance, happens to be correlated (paradox of Einstein, Podolsky, Rosen or EPR-paradox [30]). Being the direct consequence of quantum formalism, this property contradict to all our notions about causality and has no analogy within classical physics. Indeed, wave property of quantum particle has analogy in optics, stochastic behavior has analogy in statistical physics, but instantaneous correlation between noninteracting particles separated by space-like interval(up to 10 km !) seems as something like the "spooky action at distance" (as Einstein said). During last twenty years above strange property was confirmed experimentally [1,31], stimulating development such new discipline as quantum teleportation, quantum information and quantum cryptography.

We will show now that suggested topological concept  can give rather simple (although qualitative) explanation of EPR-paradox. Consider the simple example of decay of the particle with spin $0$ into two particles with spins $1/2$. Let's this will be two electrons from the ground state of helium atom. The state of these two particles after decay has the form
$$|\Psi|=2^{-1/2}(|\uparrow\rangle_{1}|\downarrow\rangle_{2}-|\downarrow\rangle_{1}|\uparrow\rangle_{2}),\eqno (23)$$ where $|\uparrow\rangle_{1}$ is the state of electron $1$ with the spin directed upwards relative to some axis (let it will be z-axis), and $|\downarrow\rangle_{2}$ is the state of electron $2$ with the spin directed downwards. The sign "minus" in Eq.(23) means that the considered state of two particles is a singlet state and not a component of triplet state (our state corresponds to the total spin $0$, but not $1$). The state (23) is an entangled state of two particles where the spin directions of each particle is not defined, but there is a quantum correlation between these directions. This correlation leads to the fact that the measurements of the spin projections of two particles appears to be correlated even after these particles are moving apart at any distance. So, quantum formalism correctly describes this phenomenon but do not give any explanation, any model of this strange behavior. The qualitative topological explanation of the paradox looks as follows (Fig.12).

\par\medskip

\begin{picture}(110,100)
\qbezier(70,70)(170,20)(270,70)
\qbezier(70,70)(175,100)(270,70)
\qbezier(121,50)(175,60)(219,50)
\put(175,95){$S=0$}
\qbezier(70,30)(170,-20)(270,30)
\put(270,40){\vector(1,0){70}}
\put(345,40){x}
\put(70,30){\line(0,1){40}}
\put(60,43){$e$}
\put(270,30){\line(0,1){40}}
\put(70,40){\circle*{5}}
\put(270,40){\circle*{5}}
\put(276,43){$e$}
\put(276,55){\vector(1,0){25}}
\put(60,55){\vector(-1,0){25}}
\put(20,40){\line(1,0){50}}
\put(75,40){\line(1,0){20}}
\put(105,40){\line(1,0){20}}
\put(135,40){\line(1,0){20}}
\put(165,40){\line(1,0){20}}
\put(195,40){\line(1,0){20}}
\put(225,40){\line(1,0){20}}
\put(255,40){\line(1,0){20}}
\put(335,5){Fig.12}
\put(286,75){Cylinder}
\end{picture}
\par\medskip

At Fig.12 two identical and not interacting electrons, propagating in opposite directions, are shown as two points of intersections of one-dimensional physical space with the their common topological defect. We took this defect as two-dimensional one (not one-dimensional) to demonstrate some symmetry effects that have no place for one-dimensional geometrical objects. We have chosen this defect as cylindrical strip because such strip is an orientable object that can be considered as a description of the state with spin $0$. Before measurement both intersection points represent according to Eq.(23) the state with spin $0$.

Suppose that the spin projection of the right electron on some direction (magnetic field {\bf H} e.g.) is measured. This measurement procedure at this point of space can change the symmetry of the whole topological defect converting it from orientable cylindrical strip into nonorientable Mobius strip (Fig.13).
\par\medskip

\begin{picture}(110,100)
\qbezier(70,70)(170,20)(270,70)
\qbezier(70,70)(175,100)(270,70)
\qbezier(121,50)(175,60)(219,50)
\put(175,95){$S=1$}
\qbezier(70,30)(170,-20)(250,30)
\put(270,40){\vector(1,0){70}}
\put(345,40){x}
\put(70,30){\line(0,1){40}}
\put(60,43){$e$}
\put(70,40){\circle*{5}}
\put(262,40){\circle*{5}}
\put(276,43){$e$}
\put(276,55){\vector(1,0){25}}
\put(286,75){Mobius strip}
\put(60,55){\vector(-1,0){25}}
\put(20,40){\line(1,0){50}}
\put(75,40){\line(1,0){20}}
\put(105,40){\line(1,0){20}}
\put(135,40){\line(1,0){20}}
\put(165,40){\line(1,0){20}}
\put(195,40){\line(1,0){20}}
\put(225,40){\line(1,0){20}}
\put(255,40){\line(1,0){20}}
\put(335,8){Fig.13}
\put(30,-2){$S=+1/2\quad or -1/2$}
\put(220,-2){$S=-1/2\quad or +1/2$}
\qbezier(267,70)(273,43)(247,28)
\put(274,65){\vector(0,1){35}}
\put(263,91){H}
\put(263,91){H}
\put(263,91){H}
\put(274,65){\vector(0,1){35}}
\put(275,65){\vector(0,1){35}}
\end{picture}

\par\medskip

So, the left intersection point instantaneously "feels" this symmetry change without any physical interaction along the physical space, and the new symmetry means realising the quantum correlation: if the electron spin projection at right equals $+1/2$, then at left it equals $-1/2$.
(Of course, "instantaneously" means "during the the time of switching on of measurement device"). In short, above schematic consideration shows how the common topological defects can serve as a channel of instantaneous informational exchange between noninteracting quantum particles separated by any distances in physical space.

EPR-paradox clearly demonstrates that apriori, before measurement, the particles had no definite values of the spin projections: quantum objects with definite physical properties appears only as a result of classical interpretation of its contact with the measurement device. Above we have demonstrated geometrical origin of such effect. Notice that some experiments with photons have shown the same properties [32], and in the next Section we will show that the notion of "electron inside the hydrogen atom" has the sense only after interaction of the atom with experimental device.

\begin{center}{\bf 8. Hydrogen atom without electron}\end{center}

The suggested geometrical interpretation of Eq. (3) can be
considered as the "kinematic" hypothesis. To be approved, it should
be verified within the dynamic problems---quantum electrodynamics,
atomic spectra, and so on. In this Section we start with the
simplest dynamic problem where the interaction is the interaction
with a given static field. Namely, we will show that the Dirac
equation for a hydrogen atom allows topological interpretation as
the equation for free Dirac field.

The Dirac equation for hydrogen atom has the form [27,28]
$$i\gamma^{\mu}(\partial_{\mu}-ieA_{\mu})\psi=m\psi.\eqno (24)$$
Here $e$ and $m$ are charge and mass of an electron, $A_{\mu}$ are
electromagnetic potentials.

It was earlier shown by Fock that the expression in (24)
$$(\partial_{\mu}-ieA_{\mu})\psi $$
can be considered as a covariant derivative of the Dirac bispinors
in the special noneuclidean space (planar Weyl space) and that
electromagnetic potentials $ieA_{\mu}$ can be considered as a
connectivities of this space [33]. Up to now, the meaning of this
result was not clear, because physical space-time does not
demonstrate any features of the Weil space in the presence of
electromagnetic field. But Fock's result acquires a physical meaning
only if we assume, on the basis of conclusions of previous Sections,
that the equation (19) is written not in the physical space, but in
an auxiliary space---universal covering space of the closed
4-manifold representing hydrogen atom.

Since the above result plays a key role, let us discuss properties
of the planar Weyl space in more detail. Geometry of this space is
specified by linear and quadratic forms [34]
$$ds^{2}=g_{ik}dx^{i}dx^{k}=\lambda (x)
(dx^{2}_{1}-dx^{2}_{2}-dx^{2}_{3}-dx^{2}_{4}),\eqno (25)$$
$$d\varphi=\varphi_{\mu}dx^{\mu},\eqno (26)$$
where $\lambda (x)$--is an arbitrary differentiable positive
function of coordinates $x_{\mu}$. This space is invariant with
respect to the scale (or gauge) transformations
$$g_{ik}^{'}=\lambda g_{ik},\quad \varphi_{i}^{'}=\varphi_{i}-
\partial \ln \lambda / \partial x_{i}.\eqno (27)$$ Therefore, a
single-valued, invariant sense has not $\varphi_{i}$ but the
quantity (scale curvature)
$$F_{ik}=\partial \varphi_{i} / \partial x_{k}-\partial
\varphi_{k} / \partial x_{i}.\eqno (28)$$ Antisymmetric tensor
$F_{ik}$ obeys equations that are analogous to the first pair of
Maxwell's equations
$$\partial_{i}F_{kl}+\partial_{k}F_{li}+\partial_{l}F_{ik}=0.$$

This analogy and the gauge invariance of  $\varphi_{i}$ (like the
gauge invariance of electromagnetic potentials) lead Weyl to the
idea that vectors $\varphi_{i}$ can be identified with the
electromagnetic potentials and that tensor $F_{ik}$ can be
identified with the tensor of electromagnetic field strengths
$$\varphi_{\mu}\equiv ieA_{\mu}, \quad A_{\mu}^{'}=
A_{\mu}-\partial_{\mu}\chi, \quad \chi = ie\ln \lambda.\eqno (29)$$
Then (like in general relativity), Weyl attempted to identify the
geometry of his space (curvature and so on) with the geometry of a
real space-time distorted by the presence of electromagnetic field
[40]. But it turned out  that  this hypothesis was contradictory to
some observable proprieties of the real physical space-time (it was
shown Einstein in the supplement to the Weyl publication [41]), and
the Weyl's results were afterwards considered as having nothing to
do with the electromagnetic field.

In contrast to Fock, we suppose that the covariant derivative in
(19) is written not in the real space-time but in the auxiliary
space--- the universal covering space of topological manifold. So,
there are no objections against the Weyl space within our
consideration. This means that we can assume that the "long
derivative" in (24) is a covariant derivative written in the Weyl
space and that the 4-potentials $ieA_{\mu}$ play the role of
connectivities in the above space. The concrete properties of the
manifold representing a hydrogen atom will be considered in
subsequent publications, but just now we can notice two important
consequences of the topological interpretation of Eq.(24).

1. It is known that connectivities of the Weyl space demonstrate the
same gage invariance as the gauge invariance of an electromagnetic
field [40]. This means that within the topological interpretation of
Eq.(24) the gage invariance of electromagnetic potentials $A_{\mu}$
is not some additional theoretical principle but is a natural
consequence of geometrical approach.

2. Geometrical interpretation of Eq.(24) for hydrogen atom does not
assume a presence of any point-like particles (electrons) inside the
atom. The wave function $\psi(x_{\mu})$ plays here the role of a
basic vector of the fundamental group representation. Coordinates
$x_{\mu}$ are coordinates of a point in the manifold universal
covering space, and this point bears no relation to some point-like
object. This fact is in agreement with suggested new paradigma. The notion of electron inside the hydrogen atom has no meaning without measurement procedure: electron can appear only as a result of interaction of the atom with measurement device (ionization and so on)

And the last remark. It seems reasonable to suppose that the same situation will
be realized within geometrical consideration of many-electrons
atoms. It is possible that the corresponding new relativistic equations (instead of nonrelativistic Schredinger equations) will turn out to be the equations for functions of only one variable $x_{\mu}$--- coordinate of the corresponding covering space, and this will give the chance to overcome known difficulties of
many-body problem of atomic physics.

\begin{center}{\bf 9.New paradigm. Perspectives}\end{center}

Within suggested model, all matter (particles, waves) is nothing more than specifically curved regions of the space itself, and acceptance of this model means refusal of modern atomic paradigm, where matter is considered as consisting of more and more small elementary particles. In the microworld there are no such classical entities as particles, waves, energies, masses and so on---only geometrical notions of curved space. These classical notions appears only as a result of the description with classical language of the contact of above pieces of curved space with measurement devices, and, of course, this result depends crucially on the measurement procedure. A priori, before measurement procedure, particles and waves do not exist, and our "macroscopic" reality is, in some way, product of above procedure.

Now---about perspectives. Suggested hypothesis should be considered as "kinematic" one, where interaction problems (except the simplest one---hydrogen atom) were not discussed. On the other hand, only application to interaction problems can demonstrate the real advantages of the proposed model. This means a necessity
of geometrization of quantized wave fields (quantum electrodynamics and so on) and geometrization of low-energy quantum physics (atomic physics and so on). We are now on this way, and we hope that the level of modern topology will be sufficient for these purposes. And, of course, the results depend on how many theoreticians will happen to be interested in the problem.

\begin{center}{\bf References}\end{center}

\noindent 1. M.B. Mensky,  Uspehi Phys. Nauk, 170, 631 (2000);   177, 415 (2007).\\
2.   A.Yu. Khrennikov (editor), Quantum Theory: Reconsideration of Foundations (Melville,New York, 2006, AIP Conf.Proc., Volume 810).\\
3.  S.l. Adler, Journal of Phys.: Conf.Ser., 67, 012014 (2007).\\
4.  Gerard 't Hooft, Journal of Phys.: Conf.Ser., 67, 012015 (2007).\\
5.  P.A.M. Dirac, Proc.of the Royal Society, A 133, 60 (1931).\\
6.  A. Einstein, Annales de la fondation Louis de Broglie, 4, 57, (1979).\\
7.  G. Loshak, La g\'{e}om\'{e}trisation de la physique, Ch.10 (Flammarion, 1994).\\
8.	 V.L. Ginzburg, About science, about myself and about others (in russian) (Moscow, Fizmatlit, 2001).\\
9.	 A. Aspect, Nature (London), 390,189 (1999).\\
10.	 W. Tittel et al., Europhys.Lett.,40, 595 (1997).\\
11.	 D. Deutsch, A. Ekert, Phys World, 11(3), 41 (1998).\\
12.	 S.Ya. Kilin, Uspehi Phys. Nauk, 169, 507 (1999).\\
13.	 J.A. Wheeler, Geometrodynamics, (New York and London, Academic Press, 1962).\\
14.	 M.B. Green, J.H. Schwaz and E. Witten, Superstring Theory (Cambridge, UK, CUP, 1987).\\
15.	O.A. Olkhov, Chem.Phys.Reports, 19(5), 1045: 19(6), 1075 (2001).\\
16.	O.A Olkhov, Chemical Physics (in russian), 21, №1, 49 (2002).\\
17.	O.A. Olkhov, Topological interpretation of of Dirac equation and geometrization of physical interactions (in russian), Preprint of Moscow Institute of Physics and Technology,  2002-1, Moscow 2002.\\
18.	O.A.Olkhov, Geometrization of some quantum mechanics formalism, Proc.Int.Conf.on
      Non-Euclidean Geometry in Modern Physics, Minsk, Belarus, 10-13 October 2006, p.145\\
19.	O.A. Olkhov, Journal of Phys.: Conf.Ser, 67, 012037 (2007); arXive: 0706.3461.\\
20.	O.A. Olkhov, Geometrization of classical wave fields, Proc.Int.Conf. "Quantum Theory: Reconsideration of Foundations," Vaxje, Sweden, 11-16 June 2007 (Melville, New York, 2008, AIP Conf.Proc., Volume 962, p.316); arXive: 0801.3746.\\
21.  J.D. Bjorken, S.D. Drell, Relativistic Quantum Mechanics
and Relativistic Quantum Fields (New York McGraw-Hill, 1964).\\
22. P. L. Rachevski, Riemannian geometry and tensor analysis (Мoscow, Nauka, 1966, \S 55.)\\
23. V.A. Jelnorovitch, Theory of spinors and its applications (Мoscow, August-Print, 2001, \S 1.3).\\
24. A.I. Achiezer, S.V. Peletminski, Fields and Fundamental
Ineractions (Kiev Naukova Dumka, 1986, Ch.1).\\
25. H.S.M. Coxeter, Introduction to Geometry (N.Y.-London John
Wiley and Sons, 1961). \\
26. B.A. Dubrovin, S.P. Novikov, A.T. Fomenko, Modern geometry
(Moscow, Nauka, 1986). \\
27. A.S. Schvartz, Quantum Field Theory and Topology (Moscow,
Nauka, 1989).\\
28. D.Gilbert, S.Kon-Fossen, Nagladnaya geometriya (rus)(Moscow, Nauka, 1981).\\
29. A.I. Achiezer, V.B. Berestetski, Quantum Electrodynamics (Moscow, Nauka, 1981).\\
30. A. Einstein, B. Podolsky, N. Rosen, Phys.Rev., 47, 777 (1935).\\
31. A. Aspect P. Grangier, G. Roger, Phys. Rev. Lett., 47, 460 (1981).\\
32. A.V. Belinsky, Uspehi Phys. Nauk, 173, 905 (2003)\\
33. V. Fock, Zs. f. Phys., 57, 261 (1929).\\
34. H. Weyl, Gravitation und Electrisitat (Berlin, Preus. Acad.Wiss., 1918).\\

\end{document}